\documentclass[%
 reprint,
showpacs,
 amsmath,amssymb,
 aps,showkeys,
prb,
floatfix,
]{revtex4-1}

\usepackage{graphicx}
\usepackage{dcolumn}
\usepackage{bm}
\usepackage[ragged]{footmisc}

\begin{document}

\preprint{APS/123-QED}

\title{Transmission properties of one dimensional  metal and left-handed  gratings}

\author{Gergely Kajt\'ar}
  \email{gergely.kajtar@stuba.sk}
\author{Peter Marko\v s}%
\affiliation{Dept. of Physics, INPE FEI STU, Ilkovi\v cova 3, 812 19 Bratislava, Slovak Republic
}%

\date{\today}

\begin{abstract}
We provide rigorous numerical calculations (using the rigorous coupled-wave analysis) of transmission spectra of various metallic and metamaterial one-dimensional gratings and identify plasmonic resonances  responsible for enhanced transmission.
We argue that the most important mechanism which influences the resonant transmission is the coupling of incident electromagnetic wave with two plasmonic waves: lengthwise plasmons, which propagates along the grating, and crosswise plasmons excited in air gaps. 
\end{abstract}

\pacs{73.20.Mf, 78.66.Bz, 81.05.Xj, 41.20.Jb}
\keywords{plasmons, metal and metamaterial  grating, enhanced transmission, RCWA}
\maketitle
    

\section{INTRODUCTION}
\label{intro}

Surface plasmon \cite{economou,maier} represents the basic eigenmode of a metallic grating. Thanks to the spatial
periodicity of the grating, surface plasmons could be excited by incident  electromagnetic (EM) wave. The interaction of these two waves (surface plasmon and incident wave) strongly influences the transmission properties of the periodic grating. 
Enhanced transmission of EM waves through periodic metallic gratings  has been  experimentally  observed in Ref.  \cite{ebbesen}. Contrary, for very thin metallic gratings, transmission is smaller due to the absorption of plasmon energy
\cite{braun}. Negative role of surface plasmons is also discussed in Ref. \cite{lalanne}.

In this paper, we calculate  the transmission of EM waves through  various one-dimensional (1D) gratings and show that  observed  resonant  frequency behavior of the transmission can be explained as an interaction of incident wave with excited plasmonic resonances. We distinguish two kinds of plasmons \cite{porto,ros}. The first one, lengthwise (LW) plasmon, which propagates along the grating surface and can be excited only when Bragg's relation between the wave vector of surface plasmon and grating period is fulfilled \cite{maier,zayats}. These LW plasmons are also responsible for the extraordinary transmission through a single aperture, if the incident surface is periodically corrugated \cite{lezec}. LW plasmon resonance appears as a sharp Fano-type peak in the spectrum. The second one, crosswise (CW) plasmon, is excited inside the gaps \cite{takakura,lee}. CW plasmon resonance appears as a broad Fabry-Perot type maximum in the spectrum. We provide quantitative estimation of resonant frequencies for both plasmons and show that they are in good agreement with rigorously calculated transmission resonances.

The Paper is organized as follows: 
In Section \ref{section2} the  transmittance of  the EM wave through a metallic grating, calculated with the use of  the rigorous coupled wave analysis (RCWA) \cite{rcwa}, is presented. In order to reveal the effect of grating geometry (grating thickness, grating period, width of gap), transmittance is calculated for different grating parameters. In Section \ref{section3}, observed resonances in the transmission spectra are interpreted as a coupling effect of incident EM wave with the surface plasmons excited on the surface of the structure or inside it. Section \ref{section4} deals with the application of the model to left-handed material gratings. Discussion of obtained results is given in Section \ref{section5}.

\section{Transmission spectra}
\label{section2}

\begin{figure}[b]
    \includegraphics[width=0.7\linewidth]{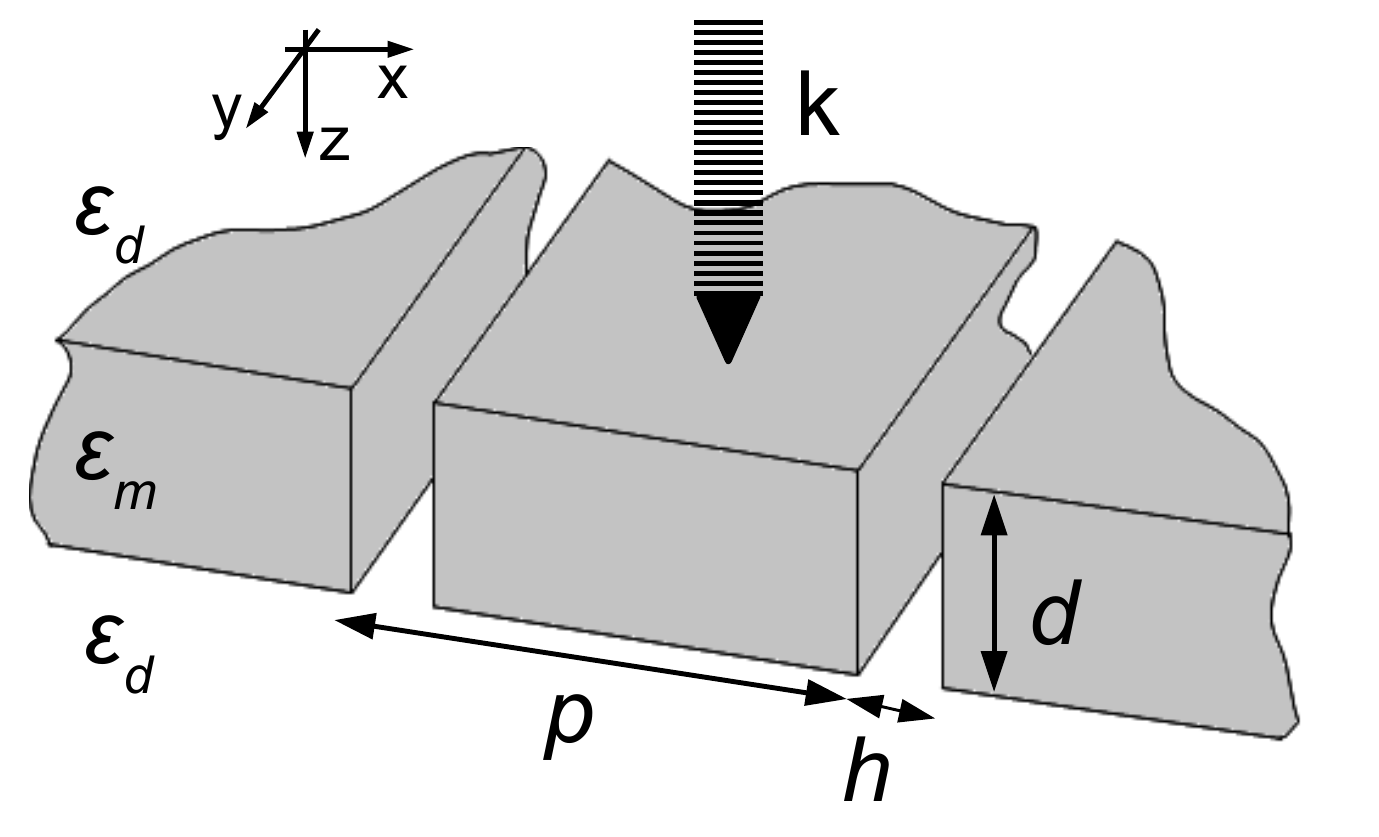}
    \caption{\label{schema1} Cross section of the 1D metallic grating.  Incident EM wave has a  wave vector $\mathbf{k}$.}
\end{figure}

\begin{figure*}
    \includegraphics[width=0.7\linewidth]{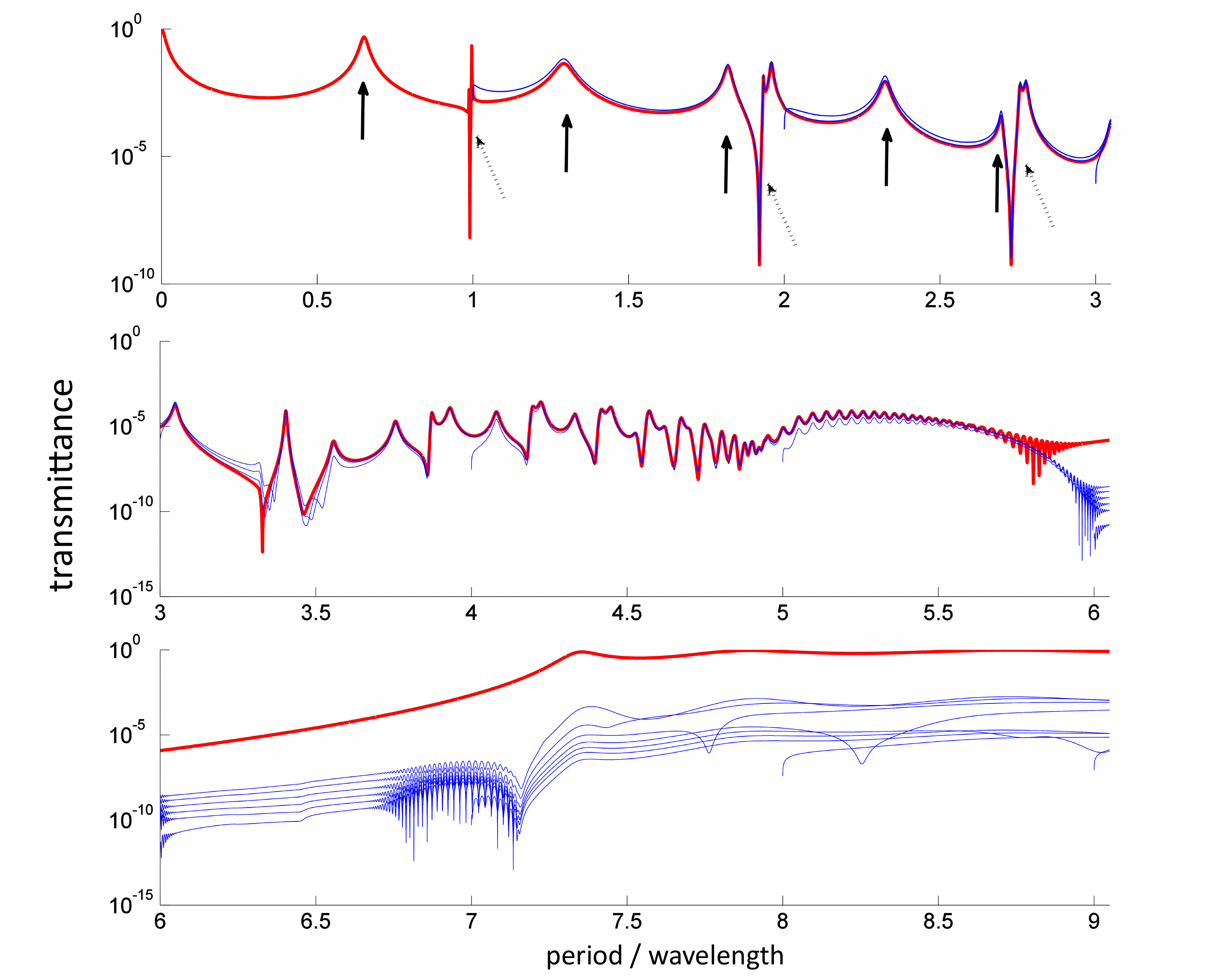}
    \caption{\label{spectrum1} (color online) Reference transmission spectrum of all diffracted orders versus the normalized frequency. Arrows indicate enhanced transmissions of  Fabry-Perot type (continuous) and Fano type (dashed arrow) resonances. Red line shows the zeroth order spectrum, blue lines represent higher order terms. Parameters of the metallic gratings  are: $p = 1000$ nm, $d = 300$ nm, $h = 10$ nm (refer to Fig. \ref{schema1}).
    Note the change of the transmission spectrum for $p/\lambda\sim 5.06$  and 7.16.  At these frequencies the real part of $\varepsilon_m$ becomes  -1 and 0, respectively.
 }   
\end{figure*}

The reference 1D metallic grating  is shown  in Fig. \ref{schema1}. The grating is uniformly extended in the $y$-direction. Its thickness in the $z$-direction is $d$ and its spatial period in the  $x$-direction is $p$.  The width of air gap is $h$.
The metallic relative permittivity is $\varepsilon_m$. The embedding medium is air with relative permittivity $\varepsilon_d=1$.   The relative permittivity of the metal is given by the  Drude's formula
\begin{equation}
\varepsilon_m\left( f \right) = 1 - \frac{f_p ^2}{f^2 + \mathrm{i}f\gamma}
\label{drude}
\end{equation}
where $f_p$ is the plasmon frequency and $\gamma$ is the absorption coefficient. For our calculations $f_p = 2147$ THz and $\gamma = 5$ THz, typical for Ag and Au \cite{metals}. 

Numerical simulations of the transmission were  realized with our own numerical program based on the RCWA. In the RCWA, the structure is divided into multiple sandwiched layers along the direction of propagation. The permittivity is a periodic function of  $x$  in each sublayer. The numerical calculation involves spatial Fourier expansion of the EM field and of the dielectric function in each sublayer of the structure.
The EM field determined by the RCWA satisfies Maxwell’s equation within each sublayer as well as the boundary conditions between adjacent layers.  The numerical accuracy is only limited by the number of orders used in the Fourier expansion.  The  typical number of Fourier modes we used is 106. If not stated otherwise, the incident electromagnetic plane wave is normal to the grating plane and the vector of magnetic field is parallel to the $z$-direction (TM polarization).


Figure \ref{spectrum1} shows the calculated transmission coefficient against the normalized frequency  $p/\lambda$. There are two types of enhanced transmission. The first one is a Fabry-Perot resonance (five lowest resonances at $p/\lambda =  0.65$, 1.30, 1.82, 2.32) are marked by arrows in Fig. \ref{spectrum1}.  The second type, Fano resonances \cite{fan}, observed at $p/\lambda = 0.99$, 1.92, 2.73, etc. are marked by dashed arrows.
Both resonances are present in all of the diffracted orders'  transmission.

To identify the physical origin of these resonances,  we  gradually change one of the parameters of the structure  and investigate how the transmission spectrum changes compared to the reference one shown in Fig. \ref{spectrum1}.

\begin{figure}[b!]
    \includegraphics[width=0.99\linewidth]{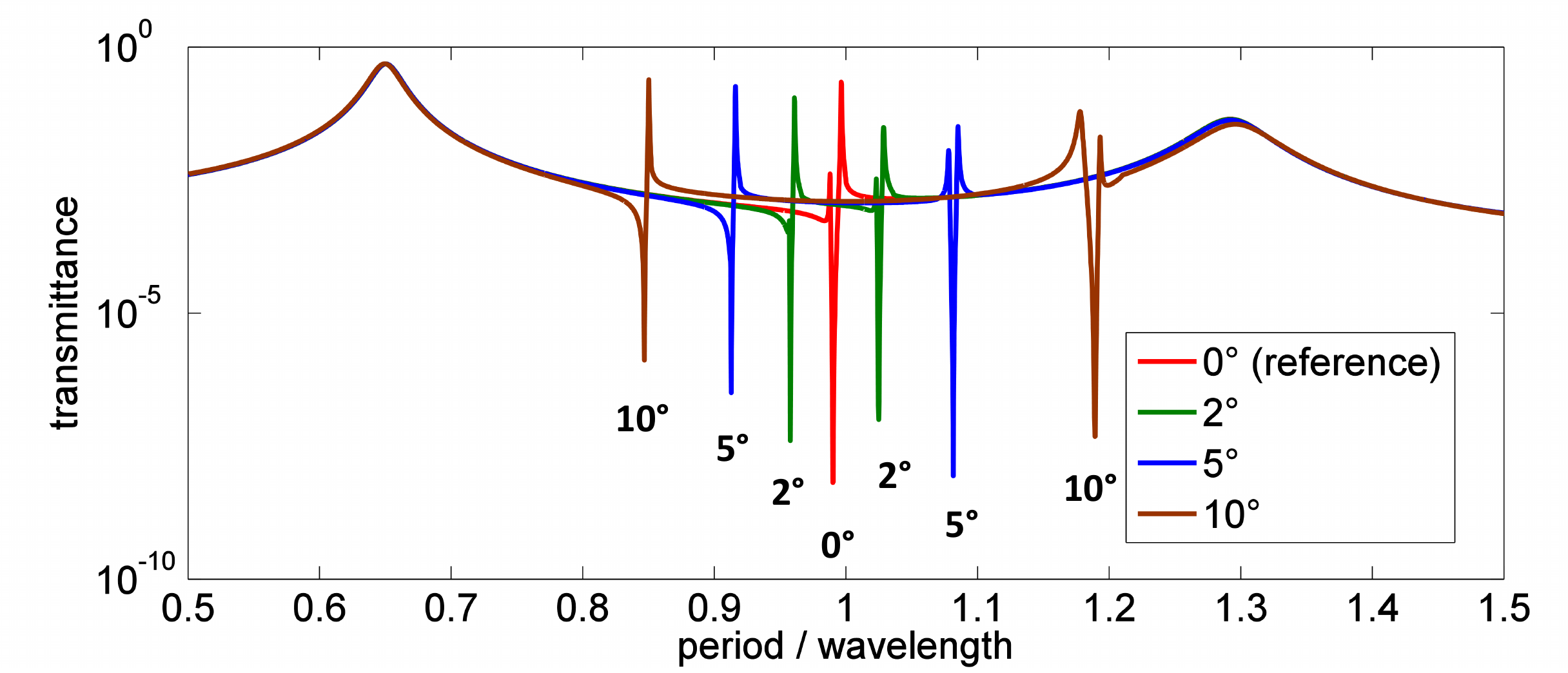}
    \caption{\label{uhol} (color online) Transmission spectrum of the zeroth diffraction order for various angles of incidence.}
\end{figure}

Figure \ref{uhol} shows how  the transmission spectra changes when  the
angle of incidence (in the $xz$ plane) increases from  $0^\circ$ (this is the reference curve from Fig. \ref{spectrum1}) to  $2^\circ$, $5^\circ$ and $10^\circ$. While  the  Fabry-Perot transmission  maxima have not changed,    Fano resonances  split to two separate frequencies which strongly depend on the incident angle. 

\begin{figure}[t]
    \includegraphics[width=0.9\linewidth]{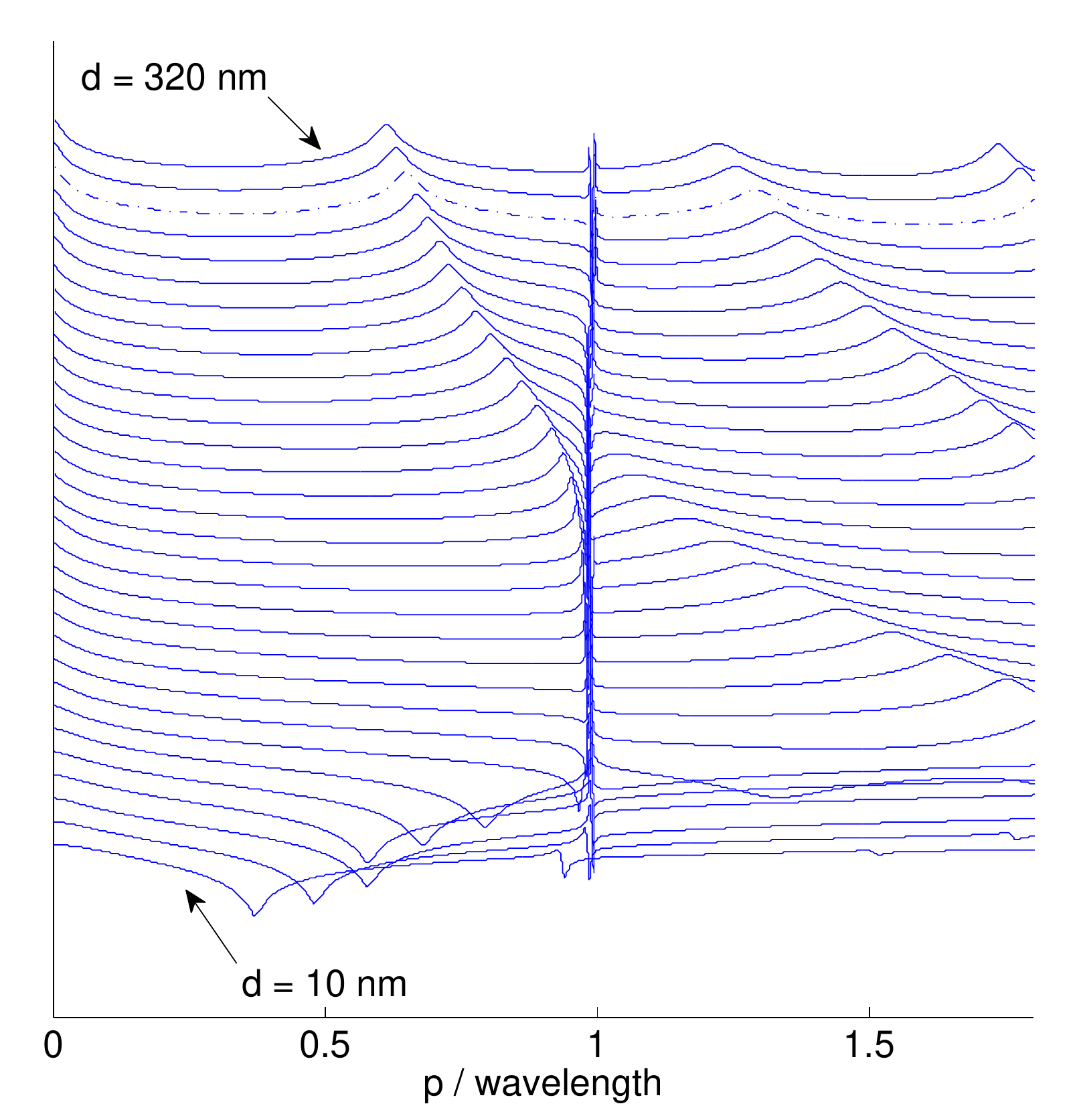}
    \caption{\label{hrubka} Transmission spectra for different grating thicknesses $d = (10,320)$ nm. Dashed line is the transmission for reference grating thickness $d = 300$ nm shown in Fig. \ref{spectrum1}.
    }
\end{figure}

Transmission spectra for various grating thicknesses $d$ are shown in Fig. \ref{hrubka}. The position of the first Fabry-Perot type maximum increases for decreasing thickness $d$. On the other hand, the position of the first Fano resonance hardly depends on $d$. It is interesting to see how the Fabry-Perot peak interacts with the Fano resonance. The ``travelling" Fabry-Perot peak actually transforms to Fano type peak and back. After further decreasing grating thickness a transmission minimum appears on interval $0 < p/\lambda < 1)$.

\begin{figure}[b]
    \includegraphics[width=0.9\linewidth]{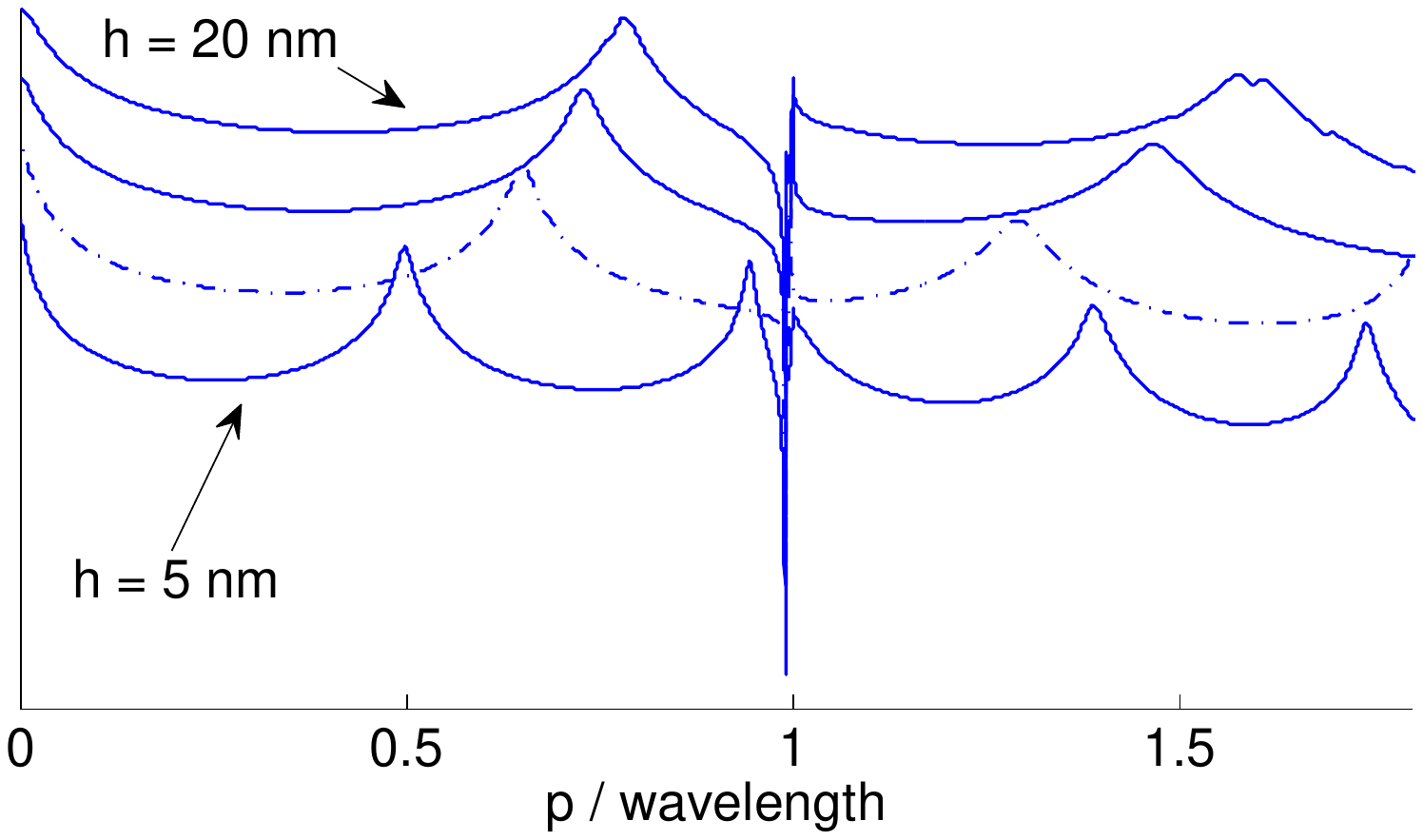}
    \caption{\label{sirka} Transmission spectra for different air gap widths $h = 5,10,15,20$ nm. Spectrum for reference air gap width $h = 10$ nm is dashed.}
\end{figure}

Similarly, a changing air gap width $h$ affects the position of Fabry-Perot type maximum (Fig. \ref{sirka}) while sharp Fano peaks remain  unaffected.

\begin{figure}[ht]
    \includegraphics[width=0.95\linewidth]{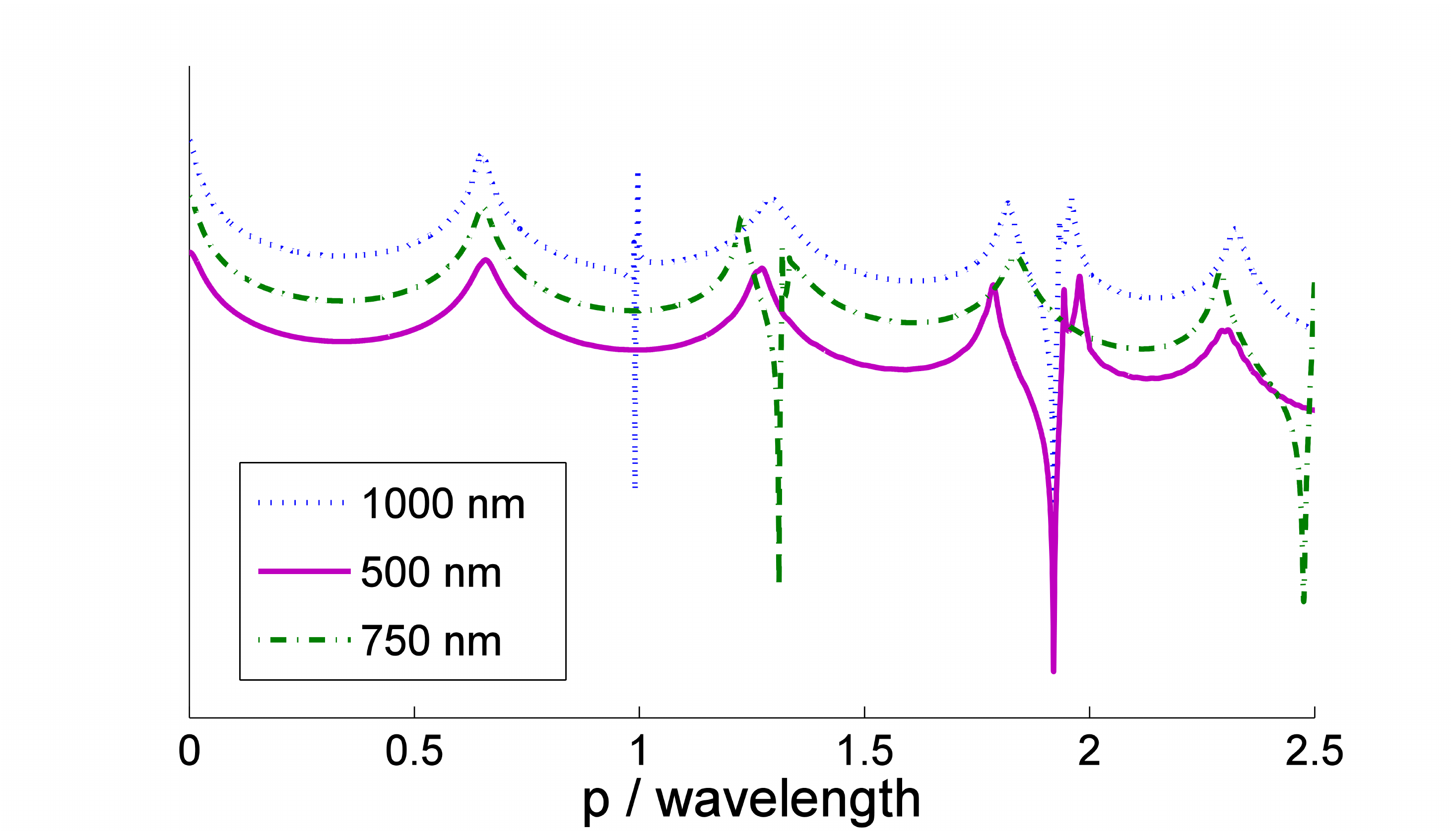}
    \caption{\label{period} Transmission spectra for different grating periods. Curves are normalized to reference period $p = 1000$ nm.}
\end{figure}

Finally, we  probe transmission against the grating period $p$. As it is seen in Fig. \ref{period}, position of Fabry-Perot maxima is left unchanged and the sharp  Fano resonances are shifted to the right for decreasing grating period $p$ and become less sharper.  

\section{Theoretical framework}
\label{section3}

\begin{figure}[b]
    \includegraphics[width=0.7\linewidth]{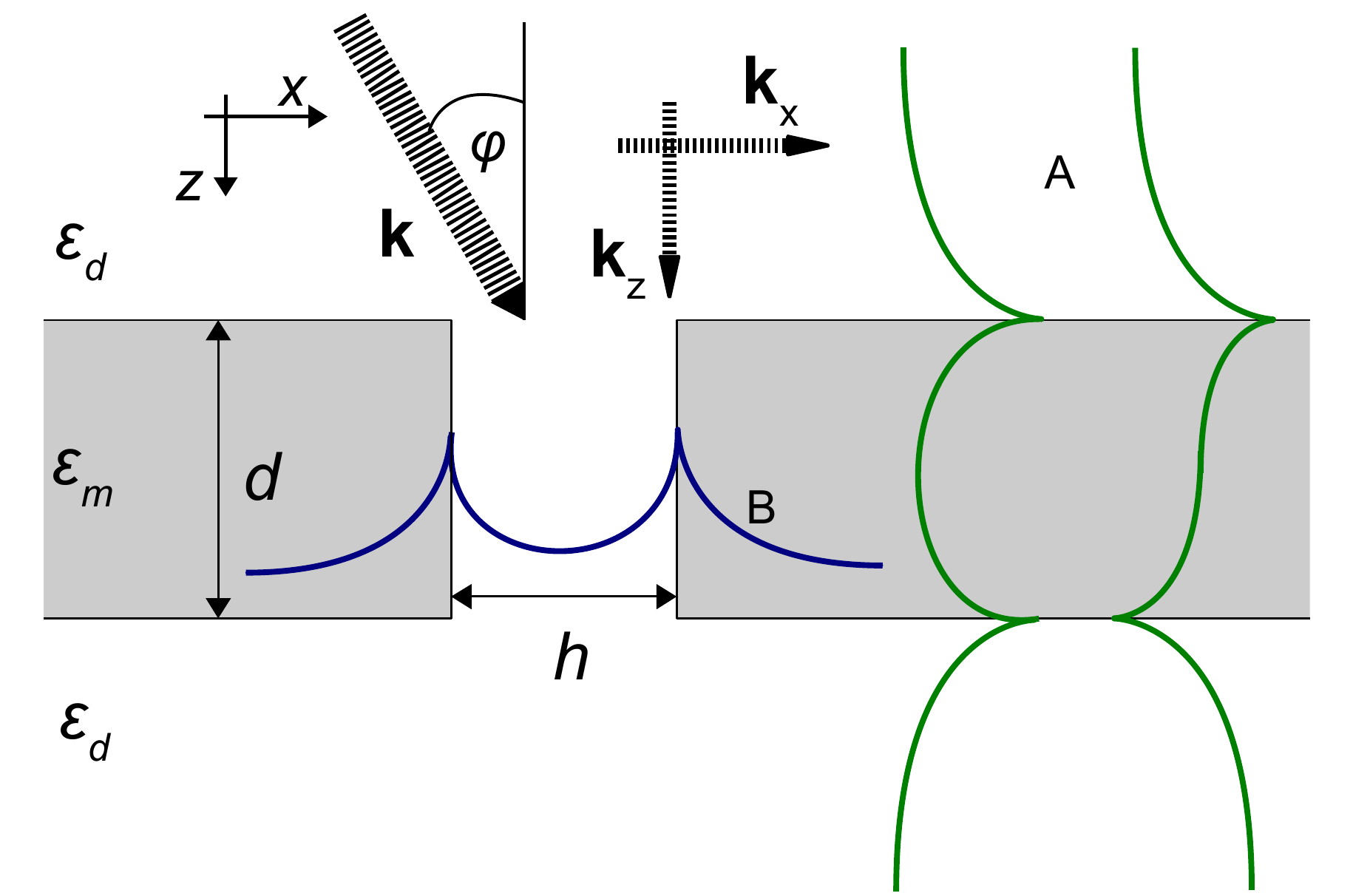}
    \caption{\label{plasmons} Three types of surface plasmons, namely lengthwise symmetric (A left) and  antisymmetric (A right) plasmon and crosswise (B) symmetric plasmon, play role in enhanced transmission.}
\end{figure}

In this section we present  an interpretation  of the origin of enhanced transmission phenomenon. We show that Fano-type sharp peaks are caused by the interaction of incident EM wave with surface plasmons propagating along the surface of the metallic grating. The excitation of these \textsl{lengthwise} (LW)  plasmons  is possible thanks to  the spatial periodicity of the grating. Similarly,  Fabry-Perot-type resonances are  due to  \textsl{crosswise} (CW)  plasmons  excited and guided along the  air gap resonators (Fig. \ref{plasmons}). 

\subsection{Lengthwise plasmons}

\def\k{\kappa}      


Consider a homogeneous metallic slab with thickness $d$ and negative relative permittivity $\varepsilon_m$ given by the real part of Eq. \ref{drude}. The metal slab is surrounded by dielectric relative permittivity $\varepsilon_d$. The dispersion relation for the  TM polarized surface waves excited on both sides of the metal-dielectrics interface is given 
\cite{economou, markos, maier} as
\begin{equation}
\frac{\varepsilon_m}{\varepsilon_d}= -\frac{\k^m _z}{\k^d _z}\tanh{\frac{\k_z ^m d}{2}}
\label{eqDisp}
\end{equation}
for the  symmetric plasmon, and 
\begin{equation}
\frac{\varepsilon_m}{\varepsilon_d}= -\frac{\k^m _z}{\k^d _z d}\coth{\frac{\k_z ^m}{2}}
\end{equation}
for the antisymmetric one. Parameters $\k_z^m$ and $\k_z ^d$  determine exponentially decreasing intensity of the EM field by the distance from the metal-air interface:
\begin{equation}
\left( \k_z ^m \right) ^2= k_x ^2 - \varepsilon_m k^2
\end{equation}
\begin{equation}
\left( \k_z ^d \right) ^2= k_x ^2 - \varepsilon_d k^2
\label{eq4}
\end{equation}
Here, $k=\omega/c$ is the wave vector of the incident wave in vacuum and $k_x$ is the $x$-component of the plasmon.
Since $\varepsilon_d$ is positive, Eq. \ref{eq4} would be satisfied only if $k_x > k$ for $\varepsilon_d = 1$. This is not achievable solely by the incident $k_x = k \sin{\varphi}$  ($\varphi$ is the angle of incidence), therefore the grating periodicity is required to modulate $k_x$:
\begin{equation}
k_x = k \sin{\varphi} + m \frac{2 \pi}{p},~~~~m=\pm 1,\pm 2,\dots
\label{eqGrating}
\end{equation}
Using Eqs. (\ref{eqDisp}-\ref{eqGrating}) we numerically obtain the positions of lengthwise plasmon resonances for different integers $m$ versus the slab thickness $d$. Results are plotted in Fig. \ref{ruppin} for normal incidence.


As it follows from Eq. \ref{eqGrating},  the position of the  LW resonance strongly depends on the angle of incidence and period $p$.  On the other hand, the dependence on the  thickness is weak, at least for sufficiently large $d$. However, air gap width has some effect on this resonance since it changes the effective relative permittivity of the metallic grating (Fig. \ref{sirka}). 

Considering these observations of numerical data from transmission spectra (Figs. \ref{uhol}, \ref{period} and \ref{hrubka}) we conclude that Fano-type sharp resonances are generated by lengthwise plasmons propagating on interfaces along the grating. Evidently, lengthwise plasmon resonance creates negative transmission effect in conjunction with Ref. \cite{lalanne}. In contrast to Ref. \cite{lalanne} we can not confirm that these resonances are closely related to diffraction orders. Although the first two sharp resonances are close to the positions of diffracted orders (because the dispersion curve of the plasmon lies very close to the light cone $\omega = ck$), higher order resonances are not connected to higher diffracted orders (see Table \ref{table}).

\subsection{Crosswise plasmons}
\label{SECcrosswise}

Crosswise plasmon resonance propagates across the grating along the interfaces of air gap (see Fig. \ref{plasmons}). Corresponding  equations are similar to that for LW plasmons:
\begin{equation}
\frac{\varepsilon_d}{\varepsilon_m}= -\frac{\k^d _x}{\k^m _x}\tanh{\frac{\k_x ^d h}{2}}
\label{eqDisp2}
\end{equation}
for the symmetric mode and
\begin{equation}
\frac{\varepsilon_d}{\varepsilon_m}= -\frac{\k^d _x}{\k^m _x}\coth{\frac{\k_x ^d h}{2}}
\end{equation}
for the  antisymmetric mode. Again, $\k_x ^m$ and $\k_x ^d$ determine exponential decrease of the EM field by the distance from the air-metal interface 
\begin{equation}
\left( \k_x ^m \right) ^2= k_z ^2 - \varepsilon_m k^2
\end{equation}
\begin{equation}
\left( \k_x ^d \right) ^2= k_z ^2 - \varepsilon_d k^2
\end{equation}
The wave vector in the $z$ direction, $k_z$, is given by a Fabry-Perot resonance condition
\begin{equation}
k_z = \frac{\pi}{d}m,~~~~m=1,2,\dots
\label{eqresonator}
\end{equation}

\begin{figure}[b]
    \includegraphics[width=0.9\linewidth]{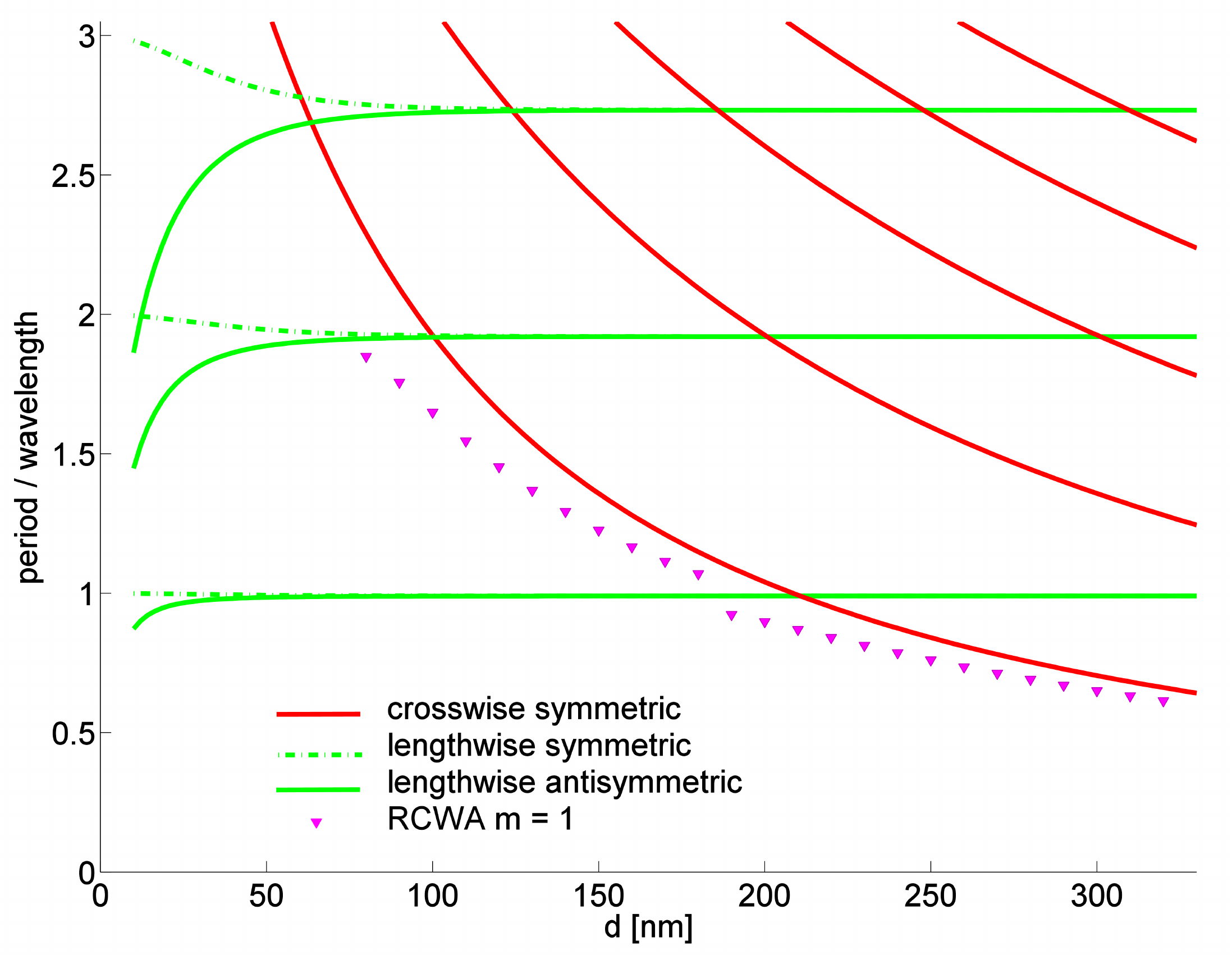}
    \caption{\label{ruppin} (color online) Plasmon resonances for different grating thicknesses $d$. There are five orders ($m=1,2,3,4,5$) showed for CW plasmons and three orders ($m = 1,2,3$) for LW plasmons, respectively.   Symbols represent numerically obtained  position of the first Fabry-Perot resonance (Fig. \ref{hrubka}) which is very close to the resonant frequency of the $m=1$ CW plasmon.}
\end{figure}

Accordingly, the wave is guided through the air gap if $k_z$ is supported by the gap-resonator's length $d$. Solving Eq. \ref{eqDisp2}-\ref{eqresonator} resonant frequencies  are obtained versus the grating thickness $d$ (Fig. \ref{ruppin}). Apparently, there is no resonance for the antisymmetric mode for the displayed interval $0< p/\lambda <3$.  

\begin{figure}
    \includegraphics[width=0.75\linewidth]{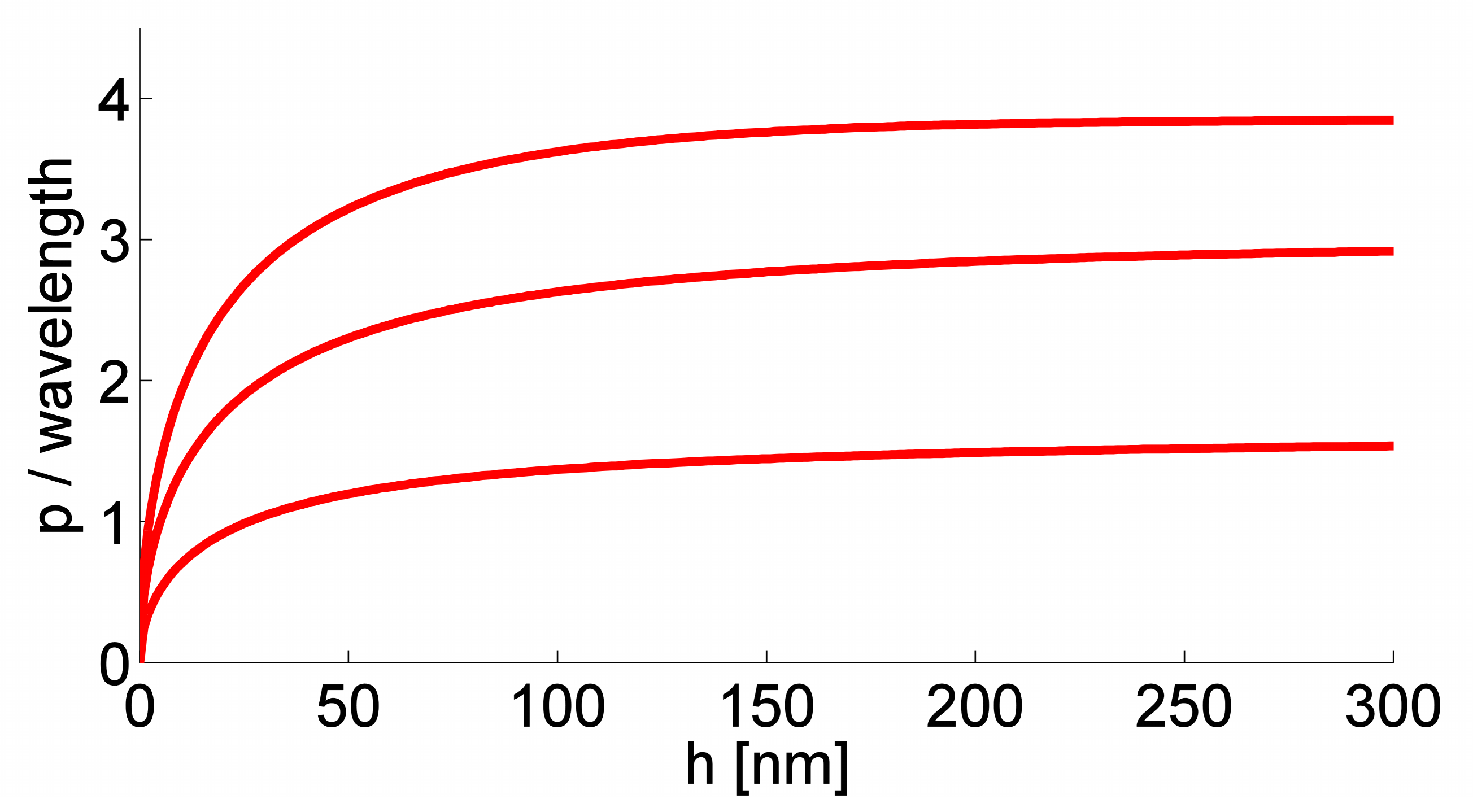}
    \caption{\label{crosswise} Symmetric crosswise plasmon resonance versus air gap width $h$ for the first three orders ($m=1,2,3$) starting from the bottom curve.}
\end{figure}

Crosswise plasmon resonance does not depend on the incident angle $\varphi$ and grating period $p$ but it depends on  the thickness $d$ and gap width $h$ (refer to Figures in Section \ref{section2} and Fig. \ref{crosswise}). This supports the idea that Fabry-Perot type resonances are  due to plasmons propagating inside air gaps.

\subsection{Identifying  plasmon resonances}

We use the presented model for the identification of numerically  calculated resonances. Calculated resonant frequencies, summarized in Tables \ref{table} and \ref{table2} confirm  that  there is a very good agreement between rigorous calculations and the model.

\begin{table}
\caption{\label{table} Comparison of calculated values using RCWA  simulation from Fig. \ref{spectrum1} and the theoretical model from Fig. \ref{ruppin}. }
\begin{ruledtabular}
\begin{tabular}{c | c c | c c}
 m & lengthwise & RCWA & crosswise & RCWA \\
\hline
1 & 0.99 & 0.99 & 0.70 & 0.65 \\
2 & 1.92 & 1.92 & 1.36 & 1.29 \\
3 & 2.73 & 2.73 & 1.93 & ~\,1.82\footnote{Can not be clearly determined because it overlaps a lengthwise plasmon resonance.\label{foot1}} \\
4 & 3.33 & 3.38 & 2.34 & 2.32 \\
5 & 3.86 & 3.85 & 2.79 & ~\,2.70\footref{foot1} 
\end{tabular}
\end{ruledtabular}
\end{table}
 
\begin{table}
\caption{\label{table2} Comparison of resonant frequencies  calculated with the  RCWA simulations  (Fig. \ref{uhol})  to the prediction of the model.}
\begin{ruledtabular}
\begin{tabular}{c c c c}
 $\varphi$ & m & RCWA & lengthwise plasmon\footnote{CW plasmon resonance is not dependent on $\varphi$.} \\
\hline
0 & 1 & 0.99 & 0.99  \\
$2^\circ$ & 1 & 1.03 & 1.02  \\
$2^\circ$ & $-1$ & 0.96 & 0.96\\
$5^\circ$ & 1 & 1.08 & 1.08 \\
$5^\circ$ & $-1$ & 0.91 & 0.91 \\
$10^\circ$ & 1 & 1.19 & 1.19 \\
$10^\circ$ & $-1$ & 0.85 & 0.85
\end{tabular}
\end{ruledtabular}
\end{table}

\begin{figure}[t!]
    \includegraphics[width=0.85\linewidth]{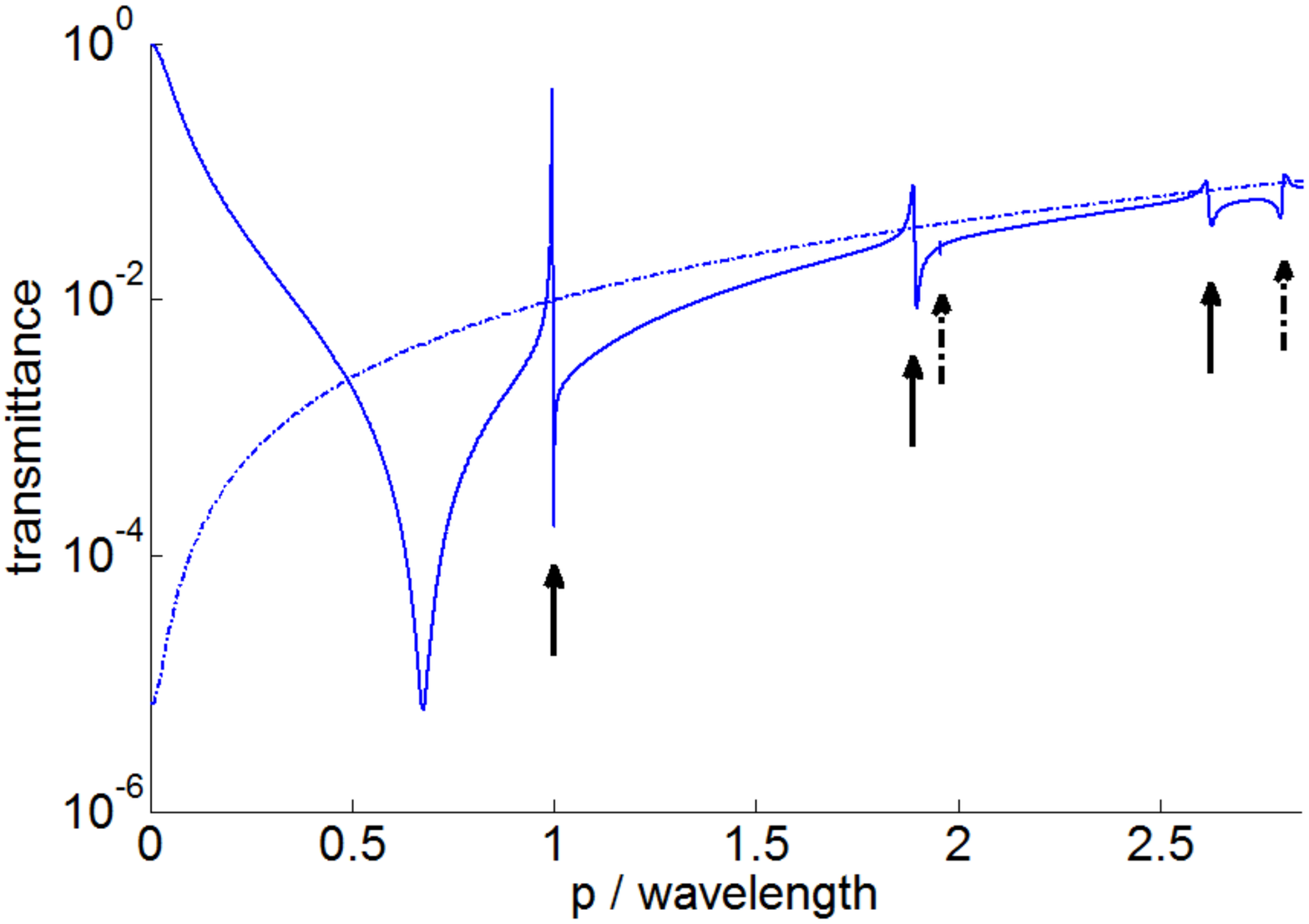}
    \caption{\label{tenkavrstva} Transmission spectrum for grating thickness $d = 39$ nm. Symmetric (dashed arrow) and antisymmetric (continuous arrow) LW plasmon resonances are visible. Dashed curve represents the transmission through homogeneous metallic slab.}
\end{figure}

For grating thickness $d = 300$~nm symmetric and antisymmetric LW plasmon resonances are barely distinguishable. To test the theory,  we calculate the transmission spectra  for thickness $d = 39$~nm (Fig. \ref{tenkavrstva}). Now, the first ($m = 1$) symmetric  and antisymmetric LW  plasmon should appear at  0.98  and  1.00, respectively.  The second 
and the third antisymmetric  and symmetric modes are clearly distinguishable at  1.86 and  1.96 ($m=2$)  and  2.58 and  2.84 ($m=3$).  This agrees with observed positions of antisymmetric LW resonances  (1.00, 1.90 and  2.62) and symmetric LW resonances (1.96 and 2.80). Apparently, the first symmetric resonance is suppressed by the first antisymmetric resonance. 

Apart from plasmonic resonances, we observe  a transmission dip at wavelength  $p/\lambda\sim 0.68$. The position of this dip depends on the width of the air gap and on the  thickness of the grating. The incoming EM wave stops to interact with the metallic grating for frequencies below this dip, thus the transmission increases to unity for the limit $\lambda \to \infty$.

\section{Left-handed material gratings}
\label{section4}

In this Section we apply the previous theory to the transmission of EM wave through periodic grating made of left-handed metamaterial (LHM).  We characterize the metamaterial by the frequency dependent permittivity and magnetic permeability using the model  described in \cite{markos}. The permittivity is given by Eq. (\ref{drude})
but with plasma frequency $f_p=10$~GHz and  absorption $\gamma = 0.01$ GHz. Frequency dependent permeability is given by the relation 
\begin{equation}\label{mu}
\mu(f) = 1 - \frac{Ff^2}{f^2 - f_0 ^2 }
\end{equation}
used for the  description of the effective  permeability of periodic array of split-ring resonators \cite{srr}.
The resonant magnetic frequency $f_0=4$~GHz and filling factor $F=0.56$.
The grating possesses a transmission band within the frequency interval 
\begin{equation}\label{in}
f_0~\le ~f~\le ~\displaystyle{\frac{f_0}{\sqrt{1-F}}}\approx 6~\textrm{GHz}.
\end{equation}
Bellow and above  this transmission band, the metamaterial behaves as a metal with frequency dependent positive magnetic permeability. 

\begin{figure}[t]
    \includegraphics[width=0.85\linewidth]{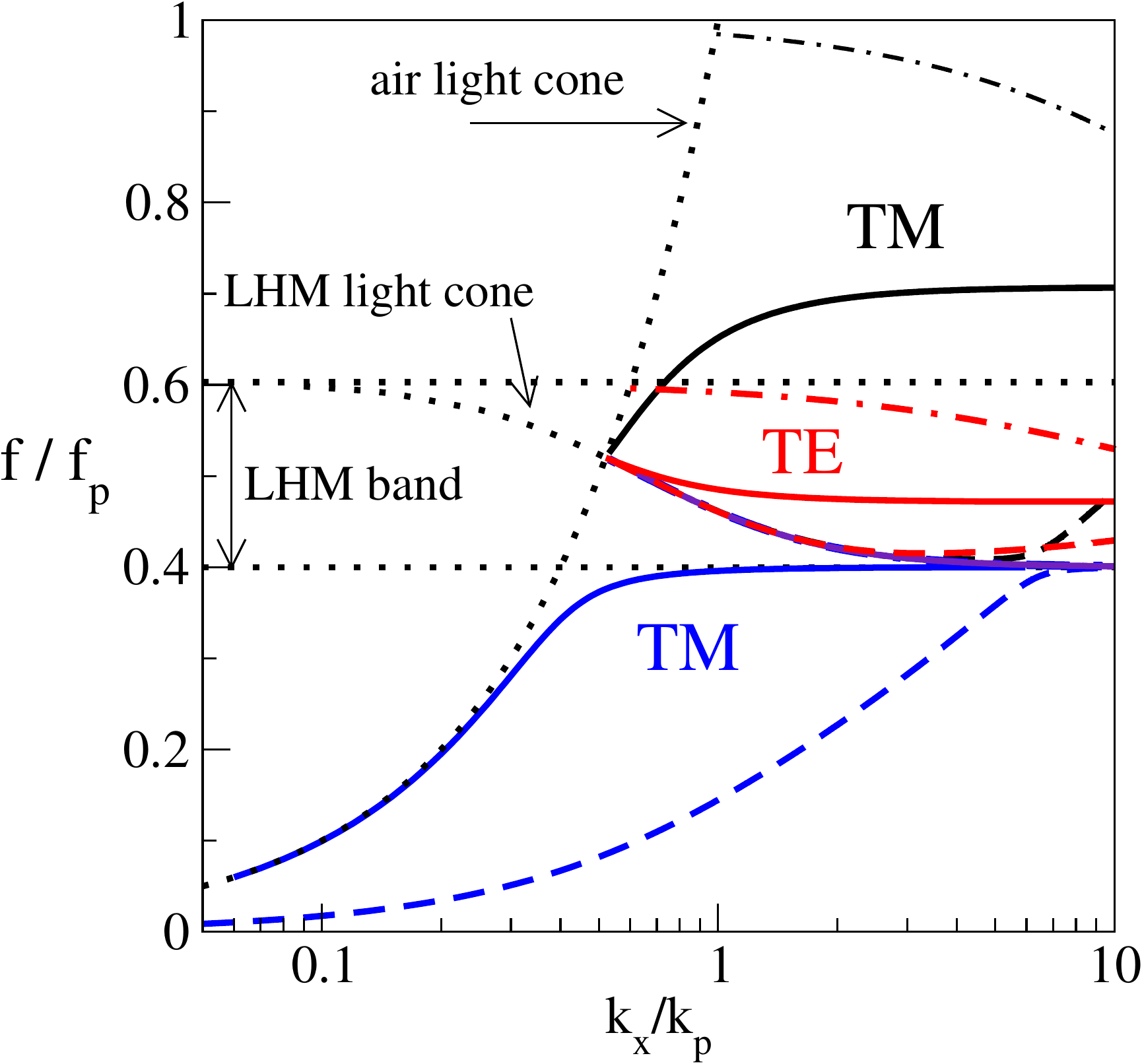}
    \caption{\label{lhm-2} (color online)  
    Thick solid lines show dispersion curves of surface plasmons propagating along the air - LHM interface.
    Note the logarithmic scale on the horizontal axis.
  For a thin air gap ($h=0.3$~mm), each plasmon   splits to a symmetric  (dashed lines) and an  antisymmetric (dot-dashed lines)  plasmon.  Note that some dispersion curves lie very close to the LHM light cone.
    }
\end{figure}

Figure \ref{lhm-2} shows the  dispersion curves of three surface plasmons propagating along the interface LHM - air
\cite{ruppin, markos}. Since the permeability of the LHM is negative on frequency interval $(0.4,0.6)$, a TE polarized surface wave might also  be excited.

\begin{figure}[h!]
    \includegraphics[width=0.95\linewidth]{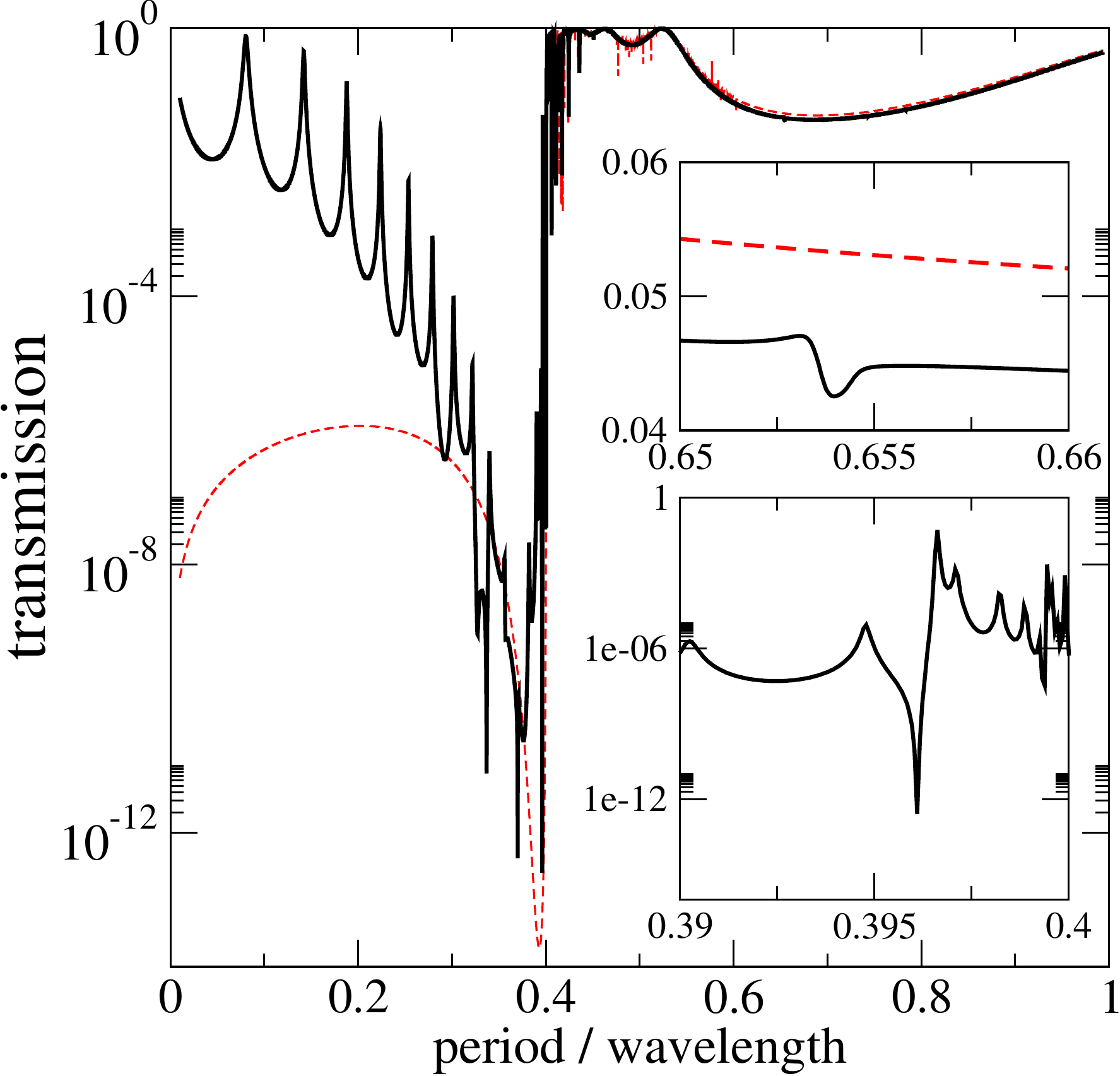}
    \caption{\label{lhm-trans} (color online) Transmission  of EM wave through the LHM grating. Black solid line: TM polarization. Red dashed line: TE polarization. Parameters of the structure are $p=3$~cm, $d= 3$~cm, air gap width $h=p/100$.
   Insets show details of the transmission in the vicinity of the lengthwise TM plasmon resonances.
   }
\end{figure}

Figure \ref{lhm-trans} presents the frequency dependent transmission through the LHM grating. As expected,
transmission is close to unity within the LHM transmission band  (Eq. \ref{in}). Nevertheless, we observed strong Fano resonances here (not clearly visible in the Fig. \ref{lhm-trans}) due to the coupling of incident EM wave with LW guided waves  \cite{shadrivov} propagating along the air-LHM-air structure in the $x$-direction and $z$-direction. For the TM modes, the dispersion curves are 

\begin{equation}
\mu(f) = \frac{k_{z}^{\rm LHM}}{\kappa_z^d}\times
\left\{
\begin{array}{l}
\tan {k_z^{\rm LHM}p}/{2}\\[3mm]
\cot {k_z^{\rm LHM}p}/{2}
\end{array}
\right.
\end{equation}
for the symmetric and antisymmetric modes, respectively. Here, 
\begin{equation}
(k_z^{\rm LHM})^2 = \varepsilon(f)\mu(f)k^2 - k_x^2 
\end{equation}
and
\begin{equation}
(\kappa_z^d)^2 = k_x^2 - \varepsilon k^2
\end{equation}
for the LHM and air, respectively. The value of $k_x$ is given by  Eq. \ref{eqGrating}.

\begin{figure}[t]
    \includegraphics[width=0.75\linewidth]{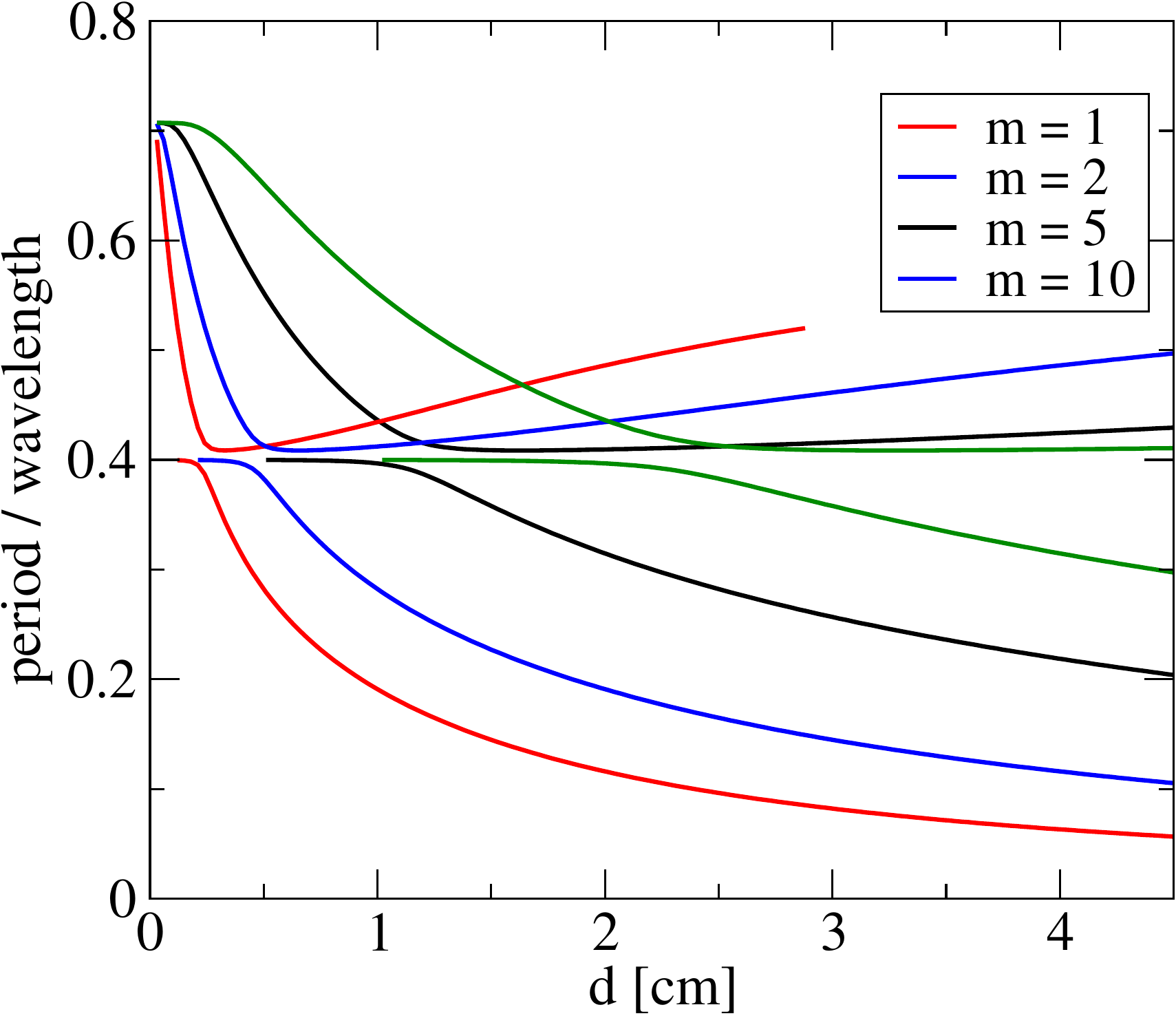}
    \caption{\label{lhm-1a} (color online) Resonant frequencies for crosswise plasmon resonances for LHM grating with  period $p=3$~cm  and air gap width  $h=p/100$ as a function of the grating thickness $d$.  The wave vector of LW plasmon is $k_z = (\pi/d)m$. CW plasmons exist also in the LHM transmission band (Eq. \ref{in}) and are responsible for fluctuations of the transmission curve.
    }
\end{figure}

On the frequency region
$f<f_0$ we observe a series of transmission maxima which could be identified as Fabry-Perot resonances based on the excitation of CW plasmons inside air gaps. Since the thickness of the grating is rather large ($d=p$), we observe large number of CW plasmons (Fig. \ref{lhm-1a}). These plasmonic resonances exist also in the transmission band (Eq. \ref{in}). 
Above the transmission band, CW plasmons cause weak oscillations of the transmission (not visible in Fig. \ref{lhm-trans}).

For the TM polarization, we identify two LW plasmon resonances at wavelengths  $p/\lambda=0.396$
and $0.65$. The latter is less pronounced since the exponential decay of plasmon is very slow $d\kappa_{\rm LHM}\sim 1$. 
We have not succeeded to identify the TE  LW plasmon resonance, expected at  $p/\lambda\approx 0.485$, because the transmission  of the TE wave  also fluctuates  in the LHM band (Eq. \ref{in}) due to excitations of CW plasmons.

\begin{table}
\caption{\label{table3} Comparison of calculated values using RCWA from Fig. \ref{lhm-trans} and the theory from Fig. \ref{lhm-1a} for LHM grating. }
\begin{ruledtabular}
\begin{tabular}{c c c}
 m & crosswise plasmon resonance & RCWA \\
\hline
1 & 0.08 & 0.08 \\
2 & 0.14 & 0.14 \\
3 & 0.19 & 0.19 \\
4 & 0.23 & 0.22 \\
5 & 0.26 & 0.25 \\
\end{tabular}
\end{ruledtabular}
\end{table}

\section{Conclusions}
\label{section5}

We have shown that enhanced transmission phenomena for periodic metallic gratings are caused by two clearly distinguishable plasmon resonances, namely lengthwise and crosswise plasmon resonances. 
Lengthwise plasmons (symmetric or antisymmetric) propagate along the two interfaces induced by the grating period. Crosswise plasmons propagate inside the air gaps.
Only symmetric  CW plasmons were observed. In the metallic gratings,  the frequency of antisymmetric CW plasmons is much higher for a narrow air gap, and  in the LHM gratings, antisymmetric  CW plasmons lie very close
to the  LHM transmission band (Fig. \ref{lhm-2}).

For completeness, we note that another excitations exist in the air gap, namely the guiding modes with both components of the wave vector, $k_x$ and $k_z$, real inside the gap.  However, similarly to 
antisymmetric CW plasmons in metallic gratings, the frequency of these excitations is not accessible  neither for metallic nor for the LHM gratings. 

Our  analysis does not  consider the role of the absorption, which is given by the small absorption parameter $\gamma$ in the frequency-dependent permittivity. Good agreement of transmission data with the prediction of the model indicates that for these parameters  the transmission
does not play significant role in the propagation of EM wave.

We have shown that plasmon resonance does not rely on diffracted orders but solely on the dispersion relation of propagating plasmons. Our rigorous calculations were used to prove the validity of a simple model of propagating plasmons. This simple model can be used to identify the type of enhanced transmission phenomenon (lengthwise or crosswise plasmon resonance) and to design various metal filters with subwavelength features which support enhanced transmission.

\medskip

This work was supported by the Slovak Research and Development Agency under the contract No. APVV-0108-11 and VEGA Agency, project No. 1/0372/13.

\def\refer#1#2#3#4#5{#1, {#2} {\bf #3}, {#4} {(#5)}}


\begin{thebibliography}{22}

\bibitem{economou} \refer{E. N. Economou}{Phys. Rev.}{182}{569}{1969};

\bibitem{maier}  S.~A.~Maier, \textsl{Plasmonics: Fundamentals and Applications} Springer Science (2002).

\bibitem{ebbesen} \refer{T. W. Ebbesen, H. J. Lezec, H.F. G
haemi, T. Thio, P. A. Wolff}{Nature}{391}{667}{1998};
\refer{H. F. Ghaemi, T. Thio, D. E. Grupp, T.~W. Ebbesen and H.~J.~Lezec}{Phys. Rev. B}{58}{6779}{1998}.

\bibitem{braun} \refer{J. Braun, B. Grompf, G. Kobiela, M. Dressel}{Phys. Rev. Lett.}{103}{203901}{2009}.

\bibitem{lalanne} \refer{Q. Cao and Ph. Lalanne}{Phys. Rev. Lett.}{88}{057403}{2002}.

\bibitem{porto} \refer{J. A. Porto, J. Garc\' ia-Vidal, J. B. Pendry}{Phys. Rev. Lett.}{83}{2845}{1999}.

\bibitem{ros} \refer{A. Roskiewicz and W. Nasalski}{ J. Phys. B: At. Mol. Opt. Phys.}{ 46}{ 025401}{2013}.

\bibitem{zayats} \refer{A.~V.~Zayats, I.~I.~Smolyaninov and A.~A.~Maradudin}{Phys. Rep.}{408}{131}{2005}.

\bibitem{lezec} \refer{H. J. Lezec et al.}{Science}{297}{820}{2002}; \refer{L. Martin-Moreno, F. J. García-Vidal, H. J. Lezec, A. Degiron, and T. W. Ebbesen}{Phys. Rev. Lett.}{90}{167401}{2003}.

\bibitem{takakura} \refer{Y.~Takakura}{Phys. Rev. Lett.}{86}{5601}{2001}.

\bibitem{lee} \refer{K.~G.~Lee and Q-Han Park}{Phys. Rev. Lett.}{98}{103902}{2005}.

\bibitem{rcwa} \refer{J. J. Hench and Z. Strako\v s}{Electron. Trans.  Num. Anal.}{31}{331}{2008}.

\bibitem{metals} \refer{I. El-Kady \textit{et al.}}{Phys Rev. B}{62}{15299}{2000}.



\bibitem{fan} \refer{Fan, Shanhui, Wonjuoo Suh, and J. D. Joannopoulos}{ J. Opt. Soc. Am. A }{20}{569–572}{2003}.

\bibitem{markos} P. Marko\v s and C. M. Soukoulis, \textit{Wave Propagation} (Princeton University Press, Princeton, 2008).

\bibitem{srr} \refer{J.~B.~Pendry et al.}{IEEE Trans. Microwave Theory Tech.}{47}{2075}{1999}.

\bibitem{engheta} N. Engheta and R.~W.~Ziolkowski (Eds.) \textsl{Metamaterials} Willey-Interscience (2006).

\bibitem{ruppin} \refer{R.~Ruppin}{Phys. Lett. A}{277}{63}{2000}; J. Phys.: Condens. Matt. {\bf 13}, 1811 (2001).

\bibitem{shadrivov} \refer{I. V. Shadrivov, A. A. Sukhorukov, and Yu. S. Kivshar}{Phys. Rev. E}{ 67}{ 057602}{2003}.



\end{thebibliography}
\end{document}